\documentclass[letterpaper,twocolumn]{article}

\usepackage[raggedright]{titlesec}

\usepackage{graphics}
\usepackage{subfigure,times,epsfig,epsf,latexsym,graphicx,multirow}
\usepackage{stfloats,amsmath,enumerate,setspace}
\usepackage{color}
\usepackage{verbatim}
\usepackage{fancybox}
\usepackage{amssymb,amsmath,amsthm}
\usepackage{tabularx}
\usepackage{algorithm,algpseudocode}
\usepackage{epstopdf}
\usepackage{array}
\usepackage{cite,url}

\titlespacing{\section}{0pt}{*2.0}{*2.0}
\titlespacing{\subsection}{0pt}{*1.8}{*1.8}

\newtheorem{definition}{Definition}

\newcommand\Mark[1]{\textsuperscript#1}

\begin{document}

\title{\Large\textbf{Designing a Human-Machine Hybrid Computing System \\ for Unstructured Data Analytics}\normalsize}

\author{
Koushik~Sinha\Mark{1}, Geetha~Manjunath\Mark{2}, Bidyut~Gupta\Mark{1} and Shahram~Rahimi\Mark{1} \vspace{5mm}\\
\begin{tabular}[t]{c@{\extracolsep{2em}}c}
\Mark{1}Dept. of Computer Science & \Mark{2}Data Analytics Research Lab \\
Southern Illinois University & Xerox Research Center India \\
Carbondale, IL, USA & Bangalore, India \\
Email: \{koushik.sinha, bidyut, rahimi\}@cs.siu.edu & Email: geetha.manjunath@xerox.com
\end{tabular}
}

\date{}
\maketitle
\thispagestyle{empty}


\begin{center}
\large\textbf{Abstract}
\end{center}
\vspace{2mm}

{\small Current machine algorithms for analysis of unstructured data typically show low accuracies due to the need for human-like intelligence. Conversely, though humans are much better than machine algorithms on analyzing unstructured data, they are unpredictable, slower and can be erroneous or even malicious as computing agents. Therefore, a hybrid platform that can intelligently orchestrate machine and human computing resources would potentially be capable of providing significantly better benefits compared to either type of computing agent in isolation. In this paper, we propose a new hybrid human-machine computing platform with integrated \textsl{service level objectives} (SLO) management for complex tasks that can be decomposed into a dependency graph where nodes represent subtasks. Initial experimental results are highly encouraging. To the best of our knowledge, ours is the first work that attempts to design such a hybrid human-machine computing platform with support for addressing the three SLO parameters of accuracy, budget and completion time.

\medskip
\noindent
\textbf{keywords:} Crowdsourcing, task scheduling, human augmented computing, service level objectives, microtask, data analytics.}



\section{Introduction} \label{intro}

With the proliferation of the Internet, mobile devices, social media and video cameras, traditional methods of data analytics  are facing disruptions. There has been a paradigm shift in data sources from the traditional format of structured data (typically arranged in rows and columns with a well-defined data model) to unstructured data where either the data does not have a pre-defined data model or is not organized in a pre-defined manner. These unstructured data comes in the form of natural language text, speech, images and videos, among others. Most state of the art automated algorithms available today do not meet the expectations of quality when it comes to analysis of these forms and volume of unstructured data.This results in irregularities and ambiguities that make it difficult to derive interpretations using traditional analytics techniques that have been designed to work with data stored in fielded form in databases or annotated (semantically tagged) in documents. The problem is even more accentuated by the volume of data that has to be dealt with, often in gigabytes or petabytes of data per day. Most studies state that today, unstructured information might account for more than $80\%$ of all data. Therefore, there is a growing requirement for human-like analytics with machine-like throughput and predictability. As an illustrative example, a company whose products are sold via millions of small retail outlets might want to gain insights from in-store video data into shopper demographics, brand/product perception and purchase behavior, in order to come up with effective sales and marketing strategies, develop new products, and optimize its supply chain. State of the art automated solutions for automatically spotting event of interests in audio-video streams perform poorly in the presence of poor lighting, clutter and crowd, ambient noise levels, multiple simultaneous conversations, and the use of multiple languages and dialects. As another example, a marketing campaign may like to perform richer qualitative analysis (theme, sentiment and intent analysis) of social media from Twitter and Facebook. Most existing automated solutions for both the above illustrative cases today can not match the quality of analysis by humans.

We therefore believe that for problems such as the above, the use of human intelligence via crowdsourcing can significantly analytics solutions. Such an approach is particularly attractive for developing economies given the large pool of human resources and high levels of under-employment. Statistics from the popular crowdsourcing platform Amazon Mechanical Turk (mTurk) reveal that the majority ($46\%$) of workers on mTurk are from developing nations, and for $30\%$ of these workers, mTurk is their main source of income \cite{mturk, ipeirotis}. Our objective is thus to design a next generation hybrid computing platform that will utilize on-demand scalable human intelligence through crowdsourcing to complement scalable machine computation on the cloud. Such a platform would theoretically be able to deliver smarter, richer and more sophisticated analytics on unstructured data. Creating such a platform poses a number of technical and software engineering challenges when we try to provide service level guarantees in terms of budget (payment for on-demand human and machine resources), completion time (humans with required skills are not always available), and accuracy (automated technology is brittle, and humans can be unreliable and error-prone).

In this paper, we present a new approach for designing such an hybrid computing platform, analogous to an operating system with machine and human processors, over which a given task workload needs to be intelligently deployed to meet the service level objectives (SLOs). Analogous to the elastic, on-demand machine computing resources enabled by the cloud, the proposed platform with its set of  API libraries for i) task definition, execution and monitoring and, ii) \textsl{connector} to existing crowdsourcing and social media platforms would enable scalable, on-demand human processors in a transparent and seamless manner. To the best of our knowledge, ours is the first work that attempts to build a computing platform that transparently uses human and machine computing agents to address the three SLO parameters of accuracy, budget and completion time deadline for solving complex, workflow based tasks.

\section{Related Work} \label{related}

A crowdsourcing platform acts as an intermediary/broker or marketplace between task \textsl{requesters} and workers for short-term microtask assignments that usually require a low degree of cognitive load and skills. While micro-tasking based crowdsourcing platforms are increasingly seeing substantial use - such as Amazon Mechanical Turk (AMT) \cite{mturk}, CrowdFlower \cite{crowdflower} and MobileWorks \cite{mobileworks} - none of them focus on automated SLO management using a hybrid man-machine computation system. AMT is the largest and most popular paid crowdsourcing platform. But, it does not provide any guarantee on either accuracy or time and the pay per microtask is fixed during \textsl{human intelligence task} (HIT) creation \cite{ipeirotis2010}. Compared to AMT, CrowdFlower provides automatic quality control through the insertion of \textsl{gold data} - data whose answers are known a priori. However, timeliness and budget adjustments are done manually by the job submitter/requester in CrowdFlower. MobileWorks uses a combination of captive workers and local managers on ground to control accuracy and time constraints. CrowdForge \cite{crowdforge} uses general purpose framework for complex, interdependent tasks \textsl{map-reduce} like abstraction to create dynamic partitioning of microtasks among workers - workers decide a task partition and once the submit their answers, their results in turn generate new subtasks for other workers. Crowdforge uses a variety of quality control methods such as voting, verification or merging items and intelligent aggregation of results using both machine algorithms as well as human workers.

Clowder \cite{weld2011} and its predecessor TurKontrol \cite{turkontrol, dai2013} use a new approach of decision-theoretic control methodology for iterative workflows (workflows where multiple passes/workers iterate over previous results) wherein each controller runs a \textsl{partially observable markov decision process} (POMDP) \cite{dai2013}. However, due to high-dimensional and continuous state space, solving a POMDP is a notoriously hard problem, thus making these approaches computationally intensive. Turkomatic \cite{kulkarni2011} is another example of iterative workflow based quality control in crowdsourcing.

The AutoMan system provides an environment where the job requester can program a confidence level for the overall computation and a budget. The AutoMan runtime system then transparently manages all details necessary for scheduling, pricing, and quality control through automatic scheduling of human tasks for each computation until it achieves the desired confidence level. The runtime system monitors, reprices, and restarts human tasks as necessary with the ability to parallely schedule the same task across multiple human workers to achieve the specified confidence level while staying under budget. The system periodically determines the minimum number of tasks necessary to meet the confidence SLO with remaining budget and spawns more tasks if required (same pay and time-out). However, AutoMan focuses only on human based computation and ignores the use of a heterogeneous computing model using machine and human agents in parallel.

In CDAS \cite{liu2012}, an analytics job is first transformed into human jobs and computer jobs, which are then processed by different modules. The human jobs are handled by the crowdsourcing engine. However, CDAS focusses only on accuracy/quality of the results. BudgetFix \cite{thanh2014} aims at crowdsourcing at minimal cost and with predictable accuracy for complex tasks that involve different types of interdependent microtasks
structured into complex workflows. BudgetFix determines the number of interdependent micro-tasks and the price to pay for each task given budget constraints. It also provides quality guarantees on the accuracy of the output of each phase of a given workflow. However, it does not consider deadline constraint and focusses only on the budget and accuracy constraints.

\section{Preliminaries} \label{prelims}

We first present a brief introduction to some of the common crowdsourcing terminology. A \textsl{crowd} refers to a group of workers willing to voluntarily do small duration and simple tasks on a crowdsourcing platform. This group is characterized by being heterogeneous and by the fact that its members do not know each other. An individual who is a member of such a crowd is known as a \textsl{crowdworker} or simply a \textsl{worker}. \textsl{Microtasking} is the process of breaking down a \textsl{task} into smaller, well defined sub-tasks known as \textsl{microtasks}. The following characteristics of a task are usually required for it to qualify as a microtask:

\begin{itemize}

\item A microtask can be performed independent of other microtasks.

\item A microtask requires human participation or intelligence and can be done in a short period of time by a human (typically ranging from a few seconds to minutes of cognitive load).

\item A microtask is either not solvable by a machine algorithm or the quality of the machine solution is unsatisfactory for the application for which the microtask was generated or would take significantly longer time than a human.

\end{itemize}

Examples of microtasks include image tagging and categorization, digitization and validation of text in images, object tagging in images, sentiment analysis of a text snippet, text classification, language translation, event detection in video, keyword spotting in audio, etc. to name a few.

\begin{figure}
\centering
\includegraphics[height=0.8in, width=2.4in]{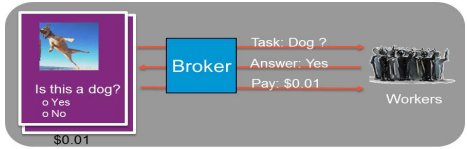}
\caption{Example of a paid microtask} \label{figure1}
\end{figure}

In this paper, we shall use the term \textsl{crowdsourcing} and \textsl{paid crowdsourcing} interchangeably for human based computing. Platforms like the AMT exhibit a list of available tasks - each task being a collection of multiple microtasks (usually tens or hundreds). A task typically includes instructions for the workers, reward per microtask and deadlines. Most tasks have two types of deadlines - one after which the task expires; the second is the time to completion before which a worker must complete her microtask to be considered for payment. Compensation per microtask is generally low, since requesters expect that work can be completed on a time scale ranging from seconds to minutes. Pay per microtask ranges from $\$0.01$ to to several dollars. As an example, on AMT, which is one of the largest paid crowdsourcing platforms, most microtasks or \textsl{human intelligence tasks} (HITs) as they are known on AMT, are priced between $\$0.01$ and $\$0.05$. A typical example of a microtask from \cite{sinha2015} is shown in Figure~\ref{figure1}. We now give a definition of \textsl{workflow} that we will use for the rest of the paper:

\begin{definition} \label{definition1}
The workflow for a given task is a dependency graph consisting of a sequence of activities, each executed by some computing entity, in order to transform raw data into useful, application specific information. The activities are represented as nodes in the dependency graph, with edges depicting data/execution dependencies/order between the activities.
\end{definition}

\section{Proposed Platform} \label{approach}

Our goal in this paper is to solve complex workflow-based tasks of the following form on a human-machine computing platform:

\textsl{Given a task $S$ that can be represented as a dependency graph consisting of nodes representing subtasks that are solvable using a human-machine agent system and the SLO metrics ($A^*$, $B^*$, $T^*$) specified by the task requester, complete $S$ with confidence/accuracy of the results being at least $A^*$, while ensuring that the total money spent is less than $B^*$ and the total time taken is less than $T^*$.}

Towards this end, we propose a hybrid computing platform that uses machine computation as well as crowd-intelligence to solve workflow-based complex analytics problems on unstructured data under a fundamental assumption that there exists a feasible solution for the specified SLOs and enough number of human workers are reachable through our platform. The most important components of our proposed platform are outlined below.

\begin{figure}
\centering
\includegraphics[height=2.1in, width=3.4in]{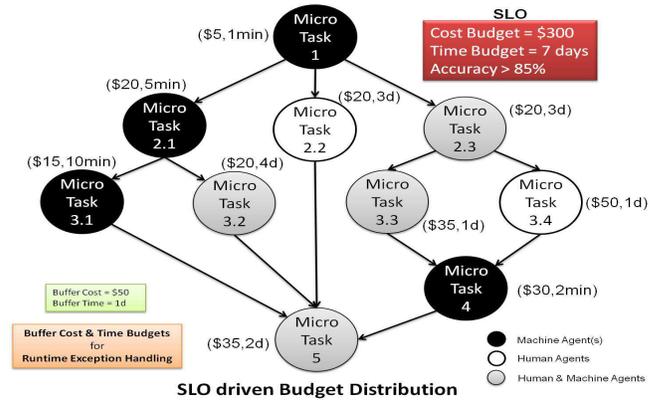}
\caption{Representative workflow description as task dependency graph} \label{figure2}
\end{figure}


\subsection{Platform Interfaces}

The platform will provide three main interface libraries for users to interact with the platform in order to create, manage, execute and monitor tasks:

\begin{enumerate}

\item \textbf{Crowd Access Layer}:  The \textsl{crowd access layer} (CAL) will be used for bringing crowd workers into our platform. The larger and more diverse the crowd, more the chances of scaling computation and meeting the SLO requirements of the tasks executing on our platform. Hence, the platform will not only provide a dedicated crowdsourcing channel for creating a private crowd for a given task, but also have connectors to tap into existing crowdsourcing platforms like the AMT and CrowdCloud. As part of the CAL, the platform will also have APIS that support a subset of the APIs of popular social media platforms (e.g., Facebook, Twitter, Instagram) so that the platform can exploit social media to recruit suitable workers for a given task.

\item \textbf{Machine Abstraction Layer}: The \textsl{machine abstraction layer} (MAL) would provide the necessary APIs that would allow platform users to plugin/register their own machine algorithms/software. To make it easy to use the platform, we intend to also provide a set of popular analytics software and APIs to access them, for text and image analytics using a \textsl{software as a service} (SaaS) model.

\item \textbf{Task Management Library}: The primary goal of the \textsl{task management library} (TML) would be to expose a \textsl{task-workflow} specification interface for hybrid tasks, whereby a complex, multi-stage task that can be decomposed into subtasks and specified as a \textsl{task-dependency graph} with subtask nodes tagged either as \textsl{machine only} or \textsl{human only} or \textsl{either} and then processed using a workflow engine. The SLOs for individual subtasks and the specific algorithms or crowd to use for a subtask node would also form a part of the task-workflow. Figure~\ref{figure2} shows a representative workflow definition where a task is decomposed into subtasks along with their individual SLOs and execution agent specifications. Once the workflow is submitted to the system, the platform would automatically schedule and manage the execution of the subtasks specified in the dependency graph with the goal of satisfying the specified SLO values. The task management library would also provide APIs that can be used for monitoring the execution progress of a submitted task.

\end{enumerate}

\subsection{Task Execution Management Engine}

The heart of the proposed platform is defined by its ability to provide service level guarantees on accuracy, time and budget while using crowd and cloud computing agents. In order to provide such guarantees while using a combination of human and machine computing agents, a \textsl{task execution management engine} is required that can provide the following two fundamental functionalities:

\begin{itemize}

\item Intelligently partition work between human and machine computing agents.

\item Provide continuous monitoring and management of task execution to guarantee SLOs of accuracy ($A^*$), budget ($B^*$) and deadline ($T^*$). For example, in the event of an exception (defined by time-out, unacceptable results submitted by a computing agents, etc.), reschedule the work to a different agent.

\end{itemize}

As a first step towards building such a task execution management engine, we consider in this paper only the problem of solving \textsl{data-parallel microtasks}, i.e., similar but independent microtasks with different inputs. Our goal is:

\textsl{Given a task set $S$ consisting of $n$ microtasks that are solvable using a human-machine agent system and the SLO metrics ($A^*$, $B^*$, $T^*$) specified by the task requester, complete $S$ with confidence/accuracy of the results being at least $A^*$, while ensuring that the total money spent is within budget $B^*$ and the total time taken is less than $T^*$.}

 A task $\mathcal{S}$ consists of $n$ independent, homogeneous microtasks each of which can be executed in parallel using either human or machine computing agents. An example would be a \textsl{sentiment analysis} task on a set of independent tweets - analyzing the sentiment of each tweet would then constitute a microtask and each tweet can be analyzed by either a human or a sentiment analysis algorithm. The $n$ microtasks are to be executed on a payment-based task execution platform. The task execution management engine uses the \textsl{crowd access layer} (CAL) or the \textsl{machine abstraction layer} (MAL) APIs to access the different types of crowd workers as well as machine algorithms for microtask execution. We assume the total allocated time interval $T^*$ (for finishing the task set $S$) to be divided into $K$ polling intervals with the polling instances being $t_0$, $t_1$, $t_2$, $\ldots$, $t_{K-1} = T^*$. We denote by $n_H(t)$ the number of microtasks assigned to humans at time instance $t$. $n_M(t) = n - n_H(t)$ is the number of microtasks assigned to machine agents at time instance $t$. Since, humans are unpredictable \cite{ipeirotis} and there is no ground truth to establish the correctness of an answer, we assume that every microtask that is assigned for human agent based execution on a crowdsourcing platform is replicated $w$ times, i.e., every human-assigned microtask is done in parallel by $w$ humans.  We denote a replicated human assignment of a microtask by $w$-task. A similar assumption for replicated assignments per machine assigned tasks can be made if required, where each of the assignments is executed using a different machine algorithm. A $w$-task is said to be in \textsl{picked} state if it has been assigned to an agent but has either not been completed or the $w$-task completion deadline has not elapsed. \textsl{done} indicates a picked $w$-task that has been returned to the system by an agent. Note that a $w$-task in \textsl{done} state does not necessarily imply that the submitted answer is correct. The correctness of a submitted answer is determined only after the microtask \textsl{result evaluation phase} in which the answers for all the done $w$-tasks are considered for correctness. A separate \textsl{result evaluation module} within the \textsl{task execution management engine} would be responsible for aggregating the results of individual $w$-tasks and arriving at a consolidated accuracy/confidence level for the microtasks completed at any given point in time. At present, we are using a simple \textsl{majority voting scheme} for result aggregation - the answer choice of the majority workers is taken as the correct answer.

 For microtasks that can be scheduled to run in parallel over either machine or human computing agents, we define an internal metric called $HM$-Ratio ($\lambda$) and distribute the microtasks workload such that the number of human tasks is $\lambda$ times that of machine tasks. The rational being that while humans can produce more accurate results for the type of tasks that we are interested in, they are also slower and more expensive as computational agents. Hence, the $\lambda$ parameter can be used to manage the accuracy SLO goal while keeping the total cost under the budget $B^*$. Unfortunately, as humans are unpredictable, there is a need for continuous feedback and control of the execution. Therefore, for dynamic control of the SLO, we have defined a second probe called the \textsl{Microtask Completion Rate} ($\rho$), which reflects the rate of completion ($\rho$) of microtasks by humans and machines. At regular polling intervals, a risk estimate of meeting $A^*$ or $T^*$ is made based on the current value of $\rho$ and appropriate corrective actions are taken subject to the budget constraint of $B^*$. Examples of such corrective actions are: changing $\lambda$ to re-allocated more microtasks to machine agents, or to more workers, or to workers with higher capability, or increasing the incentive per microtask so as to attract/employ agents with better quality/speed. These concepts have been further refined in \cite{sinha2015} and an early prototype has been built and validated through simulation with actual performance data generated from anonymous crowd workers on Amazon Mechanical Turk Machine and Hewlett Packard's Autonomy IDOL.

\section{Results}

We carried out multiple experiments using AMT to assign tasks to the crowd, and Hewlett-Packard's Autonomy IDOL \cite{autonomy} for automated analysis. For evaluation purpose, we used a set of 1000 tweets, each of which was to be categorized into six intent categories and subcategories. We experimented with three types of crowd workers - i) known, expert workers consisting of team members, ii) anonymous workers from the public crowd on AMT and, (iii) anonymous workers on AMT who had passed a short training before being allowed to work on our microtasks. A \textsl{majority voting scheme} was used for result aggregation for all worker types. For expert workers, we used 3 assignments per microtask. The 3000 microtasks were completed by the expert workers over 21 days, with $91.8\%$ of the 1000 tweets achieving majority consensus. For experiment on AMT, we used a subset of 250 randomly selected tweets from the 100 tweets dataset. Untrained workers on AMT - with both 3 and 5 assignments per tweet - categorized these 250 tweets within a day. We paid each worker $0.02$ for successfully completing a microtask assignment. In the absence of gold data, the majority voting scheme on the labeled data from the expert crowd was used to determine the correct answer for each tweet. An analysis of the results showed that for AMT crowd with 3 workers per tweet, the accuracy was only $57.2\%$ and with 5 workers per tweet it was $78\%$. For the same dataset of 250 tweets, the qualified workers generated an accuracy of $80.4\%$ but took 7 days, with 3 assignments per microtask. The performance of each type of crowd on the three SLO parameters is depicted in Figure~\ref{figure3}.

\begin{figure}
\centering
\includegraphics[height=2.7in, width=3.6in]{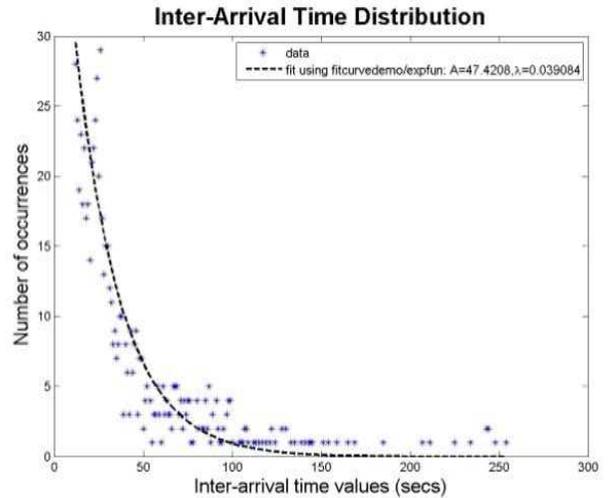}
\caption{Worker arrival pattern on AMT} \label{figure4}
\end{figure}

The results from the expert crowd were also used for training the multi-class classifier of Autonomy IDOL, which in turn was used to categorize the 250 tweets given to the AMT crowd. The corresponding accuracy for Autonomy IDOL was only $67.2\%$.

We also studied the arrival pattern of the workers on AMT to pick up our microtasks. Figure~\ref{figure4} shows the distribution of the arrival pattern of the AMT workers. It is clear from the plot that the arrival pattern of workers on AMT follows a Poisson distribution and the best fit curve had a mean arrival rate of $0.039084$.

\begin{figure*}
\centering
\includegraphics[height=2.7in, width=4.5in]{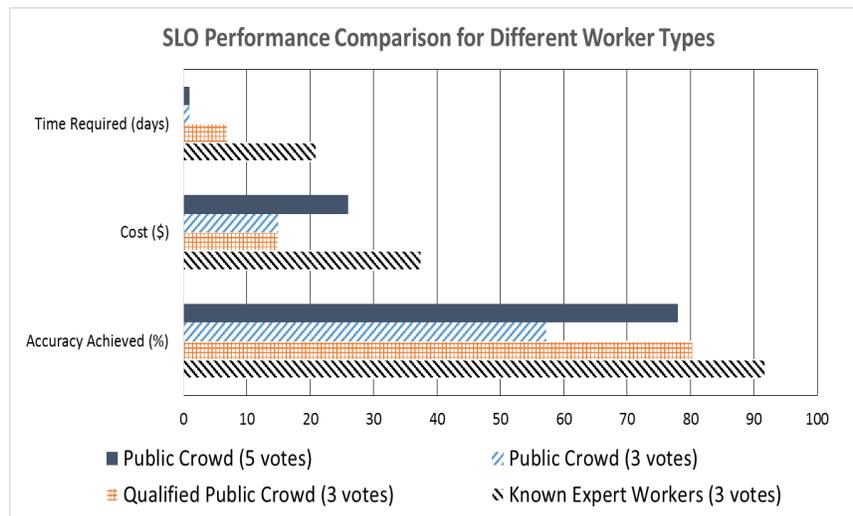}
\caption{Performance comparison of different worker classes} \label{figure3}
\end{figure*}

\section{Conclusion} \label{conclusion}

In this paper, we have proposed a new human-machine hybrid computational platform that will allow the transparent and seamless use of machine algorithms and crowdsourcing channels to solve complex, workflow-based analytics tasks on unstructured data while ensuring that the specified service level objectives of accuracy, budget and timeliness are taken into account in the task execution plan. Our initial experiments on performance data collected for anonymous crowd workers on Amazon Mechanical Turk, expert workers and machine algorithms for text analytics provide indications that such a platform would indeed be feasible and provide significant benefits for SLO driven analytics on unstructured data. We do acknowledge that more extensive experiments and development is needed to establish the complete effectiveness of our proposed platform and further work on developing a task execution management system have shown encouraging results \cite{sinha2015}. To the best of our knowledge, ours is the first work that attempts to simultaneously attempt to build a hybrid computing platform to address the three SLO parameters of accuracy, budget and deadline for data-parallel microtasks.

\end{document}